\documentclass{eas}
\usepackage{graphicx}
\usepackage{natbib}
\newcommand{\vlsr}{\ensuremath{V_\mathrm{LSR}}}
\newcommand{\K}{\ensuremath{ \textrm{ K}}}

\newcommand{\kpc}{\ensuremath{\, \mathrm{kpc}}}
\newcommand{\cm}{\ensuremath{ \textrm{ cm}}}
\newcommand{\cucm}{\ensuremath{ \cm^{-3}}}

\newcommand{\kms}{\ensuremath{ \textrm{ km s}^{-1}}}

\newcommand{\sr}{\ensuremath{ \textrm{ sr}}}

\newcommand{\ha}{\ensuremath{\mathrm{H} \alpha}}
\newcommand{\nii}{\ensuremath{\textrm{[N II]}}}
\newcommand{\sii}{\ensuremath{\textrm{[S II]}}}
\newcommand{\oiii}{\ensuremath{\textrm{[O III]}}}

\newcommand{\N}{\ensuremath{\mathrm{N}}}
\renewcommand{\H}{\ensuremath{\mathrm{H}}}

\newcommand{\arcdeg}{^{\circ}}

\begin{document}

%%-----------------------------
%%      the top matter
%%-----------------------------
\title{Ionized gas in the Smith Cloud} 
\author{Alex S. Hill}\address{Department of Astronomy, University of Wisconsin-Madison, Madison, WI USA}
\author{L. Matthew Haffner}\sameaddress{1}
\author{Ronald J. Reynolds}\sameaddress{1}
\begin{abstract}
We present WHAM observations of \ha, \nii, and \sii\ in the Smith Cloud. A map of \ha\ emission from the cloud shows ionized gas coincident with the brightest H~I emission, but nearly-as-bright \ha\ in some regions with faint H~I. The ionized mass of the cloud is at least as large as the neutral mass, $> 10^6 M_\odot$. Ionized gas in the core of the Smith Cloud has an electron temperature $6000 \K < T < 16000 \K$. The observed ratio $\nii / \ha = 0.39 \pm 0.09 $ shows that the cloud has a non-primordial nitrogen abundance, $0.1 - 1 \times$ solar.
\end{abstract}

\thanks{We thank Jay Lockman for providing the H~I data presented in \citet{lbh08}. WHAM is supported by the National Science Foundation. An American Astronomical Society International Travel Grant provided partial support for ASH to attend this meeting.}
\maketitle
%%-----------------------------
%%      your text
%%-----------------------------
%\section{Introduction}
The Smith Cloud is a high velocity cloud with a radial velocity near $+100 \kms$ located at $(l, b) = (40\arcdeg, -14\arcdeg)$. \citet{lbh08} presented an H~I survey of the cloud using the Green Bank Telescope. They indicate that the cloud is interacting with gas in the Galactic disk. The cloud has a cometary morphology with a bright core and a diffuse, trailing tail. \citet{bvc98} and \citet{pbv03} obtained \ha, \oiii\ $\lambda 5007$, and \nii\ $\lambda 6548$ Fabry-Perot spectra of two 5' fields in the core of the cloud. They detected emission from the cloud in \ha\ and \nii\ and reported an elevated $\nii / \ha$ ratio compared to the warm ionized medium.

The distance to the cloud is $12.4 \pm 1.3 \kpc$, measured using three independent techniques: stellar absorption \citep{wyw08}, kinematics of interacting gas in the Galactic disk \citep{lbh08}, and \ha\ surface brightness combined with a model of the Galactic radiation field \citep{pbv03}. This distance places the cloud 2.9~kpc below the plane and 7.6~kpc from the Galactic center with an H~I mass of $> 10^6 M_\odot$. Therefore, the cloud traces active infall onto the Galactic disk.

We present Wisconsin \ha\ Mapper (WHAM) observations of ionized gas in the cloud, including the first spectroscopic \ha\ map. Our results constrain the mass and metallicity of ionized gas in a high velocity cloud in the halo environment. An expanded version of this work is in preparation.

\begin{figure}
\includegraphics[width=13cm]{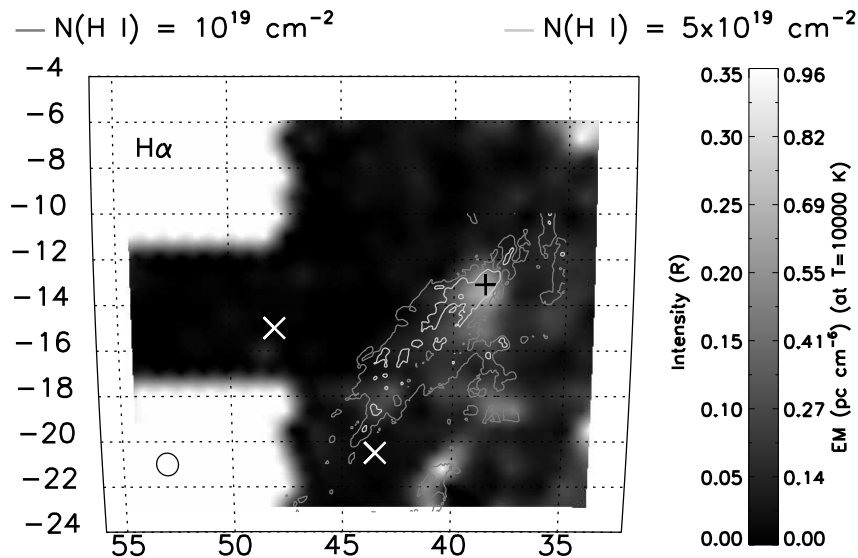}
\caption{Smoothed WHAM \ha\ maps of the Smith Cloud. Contours show H~I column densities  integrated over $+90 < \vlsr < +145$. \ha\ emission is summed over the same velocity interval. The WHAM beam is shown as a circle.}
\label{fig:smith_map}
\end{figure}

\section{Observations}

WHAM is a dual-etalon, 15~cm Fabry-Perot spectrometer coupled to a 60~cm siderostat, designed to observe faint optical emission lines from diffuse gas \citep{hrt03}. The instrument integrates all emission within its $1\arcdeg$ field of view, sacrificing spatial information to obtain spectra with a resolution of $12 \kms$ over a $200 \kms$-wide window with a sensitivity below 0.1~R in 30~s.\footnote{$1 \textrm{  Rayleigh} = 10^6 / 4 \pi \textrm{ photons s}^{-1} \cm^{-2} \sr^{-1}$}

We obtained a map of the \ha\ emission from the Smith Cloud region using WHAM with the ``block'' mapping technique described by \citet{hrt03}. We observed each $1\arcdeg$ pointing for 60~s. We integrate the observed spectrum at each pointing over the velocity range of the Smith Cloud, $+90 \kms < \vlsr < +145 \kms$; the resulting map is shown in Fig.~\ref{fig:smith_map}. This velocity interval excludes some of the emission from the cloud but avoids contamination due to the warm ionized medium gas in the foreground Sagittarius Arm.

\begin{figure}
\includegraphics[width=13cm]{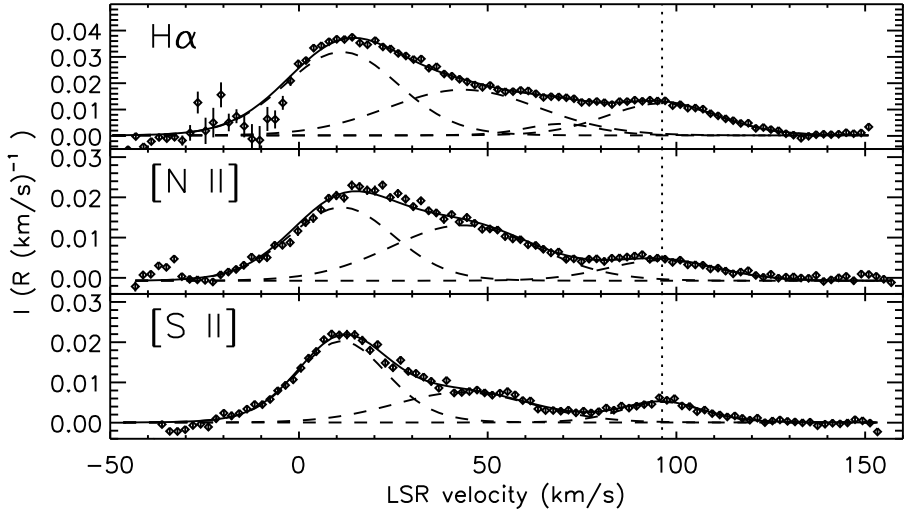}
\caption{WHAM spectra of the core of the Smith Cloud. Dashed lines show three Gaussian components used to fit the spectrum; the components correspond to local gas (near $+10 \kms$), the Sagittarius Arm ($+40 \kms$) and the Smith Cloud ($+100 \kms$). The sum of the three components is shown with a solid line.}
\label{fig:spectra}
\end{figure}

We also obtained deeper spectra towards the core of the cloud. Two-minute ``ON'' observations (black $+$ sign in Fig.~\ref{fig:smith_map}) were alternated with two-minute ``OFF'' observations (white $\times$ signs in Fig.~\ref{fig:smith_map}) for a total of six minutes on source for \ha\ and eight minutes on source each for \nii\ $\lambda 6583$ and \sii\ $\lambda 6716$. We fitted the OFF spectra of each line to an atmospheric template \citep{hrh02}, which we then subtracted from the ON spectra. Summed, atmosphere-subtracted ON spectra for each line are shown in Fig.~\ref{fig:spectra}.

We fitted the atmosphere-subtracted spectra with three Gaussians convolved with the instrument profile: one for local gas near $\vlsr=+10 \kms$, one for the Sagittarius Arm near $\vlsr = +40 \kms$, and one for the Smith Cloud near $+100 \kms$. A fourth component not clearly associated with any gas structure was required to obtain a good fit to the \ha\ data. We find a line ratio $\nii / \ha = 0.39 \pm 0.09$ in the Smith Cloud within the $1 \arcdeg$ beam of WHAM. For comparison, \citet{pbv03} found $\nii / \ha = 0.60$ within their two 5' beams.

\section{Discussion}

The \ha\ map allows an estimate of the mass of the ionized gas in the cloud. We assume that the cloud consists of uniform-density neutral hydrogen with an ionized skin of the same density. We further assume that the depth along the line-of-sight is comparable to the projected width. In the 3-square degree, high-column density core ($N_{\mathrm{H\,I}} > 5 \times 10^{19} \cm^{-2}$, outlined by lighter grey contours in Fig.~\ref{fig:smith_map}), this yields an $\H^+$ mass of $10^6 M_\odot$. Treating the entire $3 \times 1 \kpc$ cloud with $N_{\mathrm{H \, I}} > 10^{19} \cm^{-2}$ (darker grey contours in Fig.~\ref{fig:smith_map}) gas as a single, uniform-density cloud in this manner, the observed emission measure requires an H$^+$ path length (or ``skin depth'') of $> 1 \kpc$. The H~I appears patchy, suggesting that the cloud is comprised of a number of smaller clumps which each have an ionized skin. On the other hand, if we assume that the H$^+$ is evenly mixed throughout the cloud, then the density in the \ha-emitting gas is $0.015 \cucm$. The H$^+$ mass in the low-column density region of the cloud would then be $\sim 6 \times 10^5 M_\odot$. If the H$^+$ is an ionized ``skin'' with the same density as the H$^0$, the neutral clouds fill $\sim 20 \%$ of the volume of the Smith Cloud. These H$^+$ mass estimates are consistent with a previous estimate of $(0.2-1.9) \times 10^6 M_{\odot}$ (\citealt{wyw08}, based on the two pointed \ha\ measurements of \citealt{pbv03}).

The line ratio $\nii / \ha$ is sensitive primarily to the temperature and nitrogen abundance of the gas \citep{hrt99}. Because of the similar first ionization potentials of nitrogen (14.5~eV) and hydrogen (13.6~eV), we assume that $\N^+ / \N \approx \H^+ / \H$. A non-detection of \oiii\ supports our assumption that there is little $\N^{++}$ in this gas. With these assumptions, the detection of \nii\ constrains the nitrogen abundance, which scales with temperature as $\sim T^{-0.4} e^{2.2/T}$. 

Line widths of \ha\ and the relatively narrow \sii\ line constrain the temperature of the gas, assuming the ionized hydrogen and sulfur are fully mixed \citep{r85c}. From our fitted spectra, we find $T = 11000 \pm 5000 \K$ in the core. The large uncertainty is primarily due to the blending of the components due to the Smith Cloud and the Sagittarius Arm. The nitrogen abundance is $0.1-1.0 \times$ solar for the full range of possible temperatures; if we assume temperatures of $8000 - 12000 \K$ (typical of photoionized gas), the abundance is $0.2-0.5 \times$ solar. In either case, the composition of the cloud is not primordial.

\bibliography{references}

\end{document}